\documentclass[12pt,showkeys]{revtex4-1}
\usepackage[utf8]{inputenc}
\usepackage{ucs}

\usepackage{amsfonts}
\usepackage{dcolumn}% Align table columns on decimal point
\usepackage{bm}% bold math
\usepackage{graphicx}
\usepackage{hyperref}
\usepackage{subfig}
\usepackage{booktabs}
\usepackage{xcolor}
\usepackage[normalem]{ulem}
\usepackage{ucs}
\usepackage{amsfonts}
\usepackage{amssymb}
\usepackage{makeidx}
\usepackage{diagbox}
\usepackage{tabularx}
\usepackage{float} % Allows putting an [H] in \begin{figure} to specify the exact location of the figure
\usepackage{wrapfig} % Allows in-line images such as the example fish picture

\usepackage{upgreek} % para poner letras griegas sin cursiva.
\usepackage{cancel} % para tachar.
\usepackage{mathdots} % para el comando \iddots
\usepackage{mathrsfs} % para formato de letra

\begin{document}

\title{The biparametric Fisher-Rényi complexity measure and its application to the multidimensional blackbody radiation}

 \author{D. Puertas-Centeno, I. V. Toranzo, J. S. Dehesa}
 \email[]{dehesa@ugr.es}
 \affiliation{Departamento de F\'{\i}sica At\'{o}mica, Molecular y Nuclear, Universidad de Granada, Granada 18071, Spain}
 \affiliation{Instituto Carlos I de F\'{\i}sica Te\'orica y Computacional, Universidad de Granada, Granada 18071, Spain}
% \address{$^{1}$Departamento de F\'{\i}sica At\'{o}mica, Molecular y Nuclear, Universidad de Granada, Granada 18071, Spain}
% \address{$^{2}$ Instituto Carlos I de F\'{\i}sica Te\'orica y Computacional, Universidad de Granada, Granada 18071, Spain}
%  \ead{dehesa@ugr.es}

\begin{abstract}
We introduce a biparametric Fisher-R\'enyi complexity measure for general probability distributions and we discuss its properties. This notion, which is composed of two entropy-like components (the R\'enyi entropy and the biparametric Fisher information), generalizes the basic Fisher-Shannon measure and the previous complexity quantifiers of Fisher-R\'enyi type. Then, we illustrate the usefulness of this notion by carrying out a information-theoretical analysis of the spectral energy density of a $d$-dimensional blackbody at temperature $T$. It is shown that the biparametric Fisher-R\'enyi measure of this quantum system has a universal character in the sense that it does not depend on temperature nor on any physical constant (e.g., Planck constant, speed of light, Boltzmann constant), but only on the space dimensionality $d$. Moreover, it decreases when $d$ is increasing, but exhibits a non trivial behavior for a fixed $d$ and a varying parameter, which somehow brings up a non standard structure of the blackbody $d$-dimensional density distribution. 
\end{abstract}

%

\maketitle

\section{Introduction}

General quantification of complexity attributed to many-body systems, the task which is closely connected with evolution from order to disorder, is among the most important scientific challenges in the theory of complex systems \cite{badii,sen2012,seitz}. The fundamental issue is to find one quantifier which is able to capture the intuitive idea that complexity lies between perfect order and perfect disorder. Most probably this idea cannot be formalized by a single complexity quantifier because of the so many facets of the term complexity. Based on Information Theory and Density Functional Theory, various computable and operationally meaningful density-dependent measures have been proposed: the entropy and complexity measures of the one-body probability density of the system. The former ones (mainly the Fisher information and the Shannon entropy, and their generalizations like Rényi and Tsallis entropies) capture a single macroscopic facet of the internal disorder of the system. The latter ones capture two or more macroscopic facets of the quantum probability density which characterize the system, the most relevant ones being, up until now, the complexity measures of Crámer-Rao \cite{dehesa_1,antolin_ijqc09}, Fisher-Shannon \cite{romera_1,angulo_pla08} and LMC (López-Ruiz-Mancini-Calvet) \cite{lopez95,catalan_pre02,anteneodo}, which are composed by two entropic factors. These three basic measures, which are dimensionless, have been shown to satisfy a number of interesting properties: bounded from below by unity, invariant under translation and scaling transformation, and monotonicity (see e.g. \cite{guerrero,rudnicki16}). Recently, they have been generalized in various directions such as the measures of Fisher-Rényi \cite{romera08,antolin09,romera09,antolin_ijqc09,rudnicki} and LMC-Rényi  \cite{pipek,lopez,lopezr,sanchez-moreno} types.\\

The aim of this article is twofold. First, we introduce a novel class of biparametric measures of complexities (namely, the generalized Fisher-Rényi measures) for continuous probability densities, which generalizes the basic Fisher-Shannon and its extensions of Fisher-Rényi type. Second, we illustrate the utility of these novel complexity quantifiers by applying them to the generalized Planck radiation law (see below). Beyond the temperature, we will focus on the dependence of these quantifiers on the space dimensionality $d$ and the complexity parameters.

The cosmic microwave, neutrino and gravitational backgrounds (cmb, cnb and cgb, respectively) give information about the universe at different times after the big bang. The cnb and cgb have been claimed to give information at one minute after the big bang and during the big bang, respectively, and to have been seen in recent experiments with controversial results, still under a careful examination \cite{amaro-seoane,ade,follin} (see \cite{faessler} for a brief summary). The cmb, originated at around 380,000 years after the big bang at a temperature of around 3000 Kelvin, is the only cosmic radiation background which is well established. It was first detected in 1964 by the antennae manipulation works of Penzias and Wilson \cite{penzias}, and later confirmed by satellite observations in a very detailed way \cite{boggess,bennett,ade2}. It is known that the frequency distribution of the cmb which presently bathes our (three-dimensional) universe follows the Planck’s blackbody radiation law given by the (unnormalized) spectral density $\rho_{T}^{(3)}(\nu) = \frac{8 \pi h}{c^{3} } \nu^{3} (e^{\frac{h \nu }{k_{B} T}}-1)^{-1}$ at the temperature $T_0 = 2.7255(6)$ Kelvin, where $h$ and $k_B$ denote the Planck and Boltzmann constants, respectively.\\

In the last few years there is an increasingly strong interest in the analysis of the quantum effects of the space dimensionality in the blackbody radiation \cite{cardoso,ramos1,alnes,lehoucq,stewart,nozari,zeng,ramos2} and, in general, for natural systems and phenomena of different types in various fields from high energy physics and condensed matter to quantum information and computation (see e.g. \cite{acharyya,krenn,march,dehesa2,aptekarev,toranzo16,rybin,salen} and the monographs \cite{herschbach,dong,weinberg,dehesa1}). This is not surprising because of the fundamental role that the spatial dimensionality plays in the solutions of the associated wave equations \cite{herschbach,dong}. In the present work we adopt an information-theoretical approach to investigate the complexity effects of the spatial dimensionality in the spectral energy density per unit of frequency of a blackbody at temperature $T$ which has been found \cite{cardoso,alnes,ramos1} to be given in a $d$-dimensional space by the generalized Planck radiation law
\begin{equation}
\label{eq:GPRL}
%\rho_{T}^{(d)}(\nu) = \frac{2 (d-1) h \left(\frac{\sqrt{\pi }}{c}\right)^d}{\Gamma \left(\frac{d}{2}\right) }\frac{\nu^{d}}{e^{\frac{h \nu }{k_{B} T}}-1} 
\rho_{T}^{(d)}(\nu) = \frac{1}{\Gamma(d+1)\,\zeta(d+1) }\left(\frac{h}{k_{B}T}\right)^{d+1}\frac{\nu^{d}}{e^{\frac{h \nu }{k_{B} T}}-1},
\end{equation}
(normalized to unity), where $\Gamma(x)$ and $\zeta (x)$ denote the Euler gamma function and the Riemann zeta function \cite{olver}, respectively. This investigation will be done by means of a novel class of biparametric measures of complexity of Fisher-Rényi type which allows us to go further beyond the 2014-dated work \cite{toranzo14} based on the entropy-like measures of $\rho_{T}^{(d)}(\nu)$ and the three basic two-component complexity measures of Crámer-Rao, Fisher-Shannon and LMC types. \\

The structure of the work is the following. First, we define and explore the main properties of the biparametric complexities and their entropy-like components (Rényi entropy, biparametric Fisher information) of a general continuous one-dimensional probability distribution. Second, we determine and discuss the values of the previous entropy-like and complexity measures for the spectral density $\rho_{T}^{(d)}(\nu)$ which characterizes the multidimensional blackbody distribution. Then, we numerically discuss the dependence of these blackbody spectral quantifiers on the universe dimensionality and the complexity parameters. Let us advance that the resulting blackbody biparametric complexities are mathematical constants (i.e. they are dimensionless), independent of the temperature $T$ and of the physical constants (Planck's constant, speed of light and Boltzmann's constant), so that they only depend on the spatial dimensionality. Finally, some concluding remarks are given.

\section{The biparametric Fisher-Rényi complexity measure of a general density}

In this Section we define and discuss the meaning of a class of biparametric complexity measures of Fisher-Rényi type for a one-dimensional continuous probability distribution $\rho(x)$, $x \in \Delta \subseteq \mathbb{R}$. Heretoforth we assume that the density is normalized to unity, so that $\int_{\Delta} \rho(x) dx  =  1$. \\

First, we define the biparametric Fisher-Rényi complexity measure  of the density $\rho(x)$ as     
        \begin{equation}
        \label{eq:def1}
   C^{(p,\lambda)}_{FR}[\rho] =\mathcal K_{FR}(p,\lambda) \,\,\phi_{p,\lambda}[\rho] \times N_{\lambda}[\rho],
    \end{equation}
with $p^{-1} + q^{-1} = 1$, $\lambda > (p+1)^{-1}$ and where  the symbols $\mathcal K_{FR}(p,\lambda)$, $\phi_{p,\lambda}[\rho]$ and $N_{\lambda}[\rho]$ denote a normalization factor, the biparametric Fisher information \cite{lutwak} and the Rényi entropy power, $N_{\lambda}[\rho] = \exp\left(R_{\lambda}[\rho] \right)$, respectively. Moreover, the symbol $R_{\lambda}[\rho]$ denotes the Rényi entropy of order $\lambda$ defined \cite{renyi_70} as
\begin{equation}
\label{eq:renyi_entrop}
R_{\lambda}[\rho]=\frac{1}{1-\lambda}\ln\left(\int_{\Delta}[\rho(x)]^{\lambda}dx\right),
\end{equation}
with $\lambda>0$ and $\lambda\neq1$.This quantity is known to quantify various $\lambda$-dependent aspects of the spreading of the density $\rho(x)$ all over its support $\Delta$. In particular, when $\lambda \to  1$ the Rényi entropy tends to the Shannon entropy $S[\rho]$, which measures the total spreading of  $\rho(x)$. On the other hand, the biparametric Fisher information $F_{p,\lambda}[\rho]$ \cite{lutwak} is given by
\begin{equation}
  \label{eq:genfish}
	\phi_{p,\lambda}[\rho] 
									= \left(\int_{\Delta} \left|[\rho(x)]^{\lambda-2}\rho'(x)\right|^{q}\rho(x)\, dx\right)^{\frac{1}{q\lambda}} ,
  \end{equation}
 with $\frac{1}{p}+\frac{1}{q} =1$, $p\in[1,\infty)$ and $\lambda\in\mathbb{R}$. 
  Note that for the values $(p,\lambda)=(2,1)$ the square of this generalized measure reduces to the standard Fisher information, i.e., $\phi_{2,1}[\rho]^{2} = F[\rho] = \int_{\Delta}\frac{|\rho'(x)|^{2}}{\rho(x)}\,dx$. So, while $F[\rho]$ quantifies the gradient content of $\rho(x)$, the generalized Fisher information $\phi_{p,\lambda}[\rho]$ with $p\neq 2$ and $\lambda \neq 1$ measures the $(p,\lambda)$-dependent aspects of the density fluctuations other than the gradient content. 
  \noindent

The normalization factor, $\mathcal K_{FR}(p,\lambda)$, in Eq. (\ref{eq:def1}) is given by 
     \begin{eqnarray}\label{eq:KFR}
   \mathcal K_{FR}(p,\lambda) &=& (\phi_{p,\lambda}[G]N_\lambda[G])^{-1}  \nonumber\\
   &=&  \left[\frac{\lambda^\frac1q}{p^\frac1p}\,a_{p,\lambda}\,e_\lambda\left(\frac{-1}{p\lambda}\right)^{\frac{\lambda-1}{p}+1}\right]^\frac1\lambda\nonumber \\
   &=& a_{p,\lambda}^{\frac1\lambda} \left(\frac{(p\lambda+\lambda-1)^{\frac{q\lambda-\lambda+1}q}}{p\lambda^\lambda}\right)^{\frac1{\lambda-\lambda^2}},
    \end{eqnarray}
where $G$ denotes the generalized Gaussian distribution $G(x)$ \cite{lutwak} given by 
\begin{equation}
 G(x)=a_{p,\lambda}\, e_\lambda(|x|^p)^{-1}
\end{equation}
where $e_\lambda(x)$ denotes the modified $q$-exponential function \cite{borges_physA}:
\begin{equation}
e_\lambda(x)=(1+(1-\lambda)x)_ +^{\frac1{1-\lambda}},
\end{equation}
which for $\lambda \to 1$ reduces to the standard exponential one, $e_{1}(x)\equiv e^{x}$.
For $p \in (0,\infty)$, $\lambda >1-p$ and with the notation $t_{+} = \max\{t,0\}$ for any real $t$, the constant $a_{p,\lambda}$ is given by
 \begin{equation}
 \label{eq:norm_gauss}
a_{p,\lambda} = \left\{\begin{array}{ll}
   \frac{p(1-\lambda)^{1/p}}{2B\left(\frac{1}{p},\frac{1}{1-\lambda}-\frac{1}{p}\right)} & if\,\,\lambda < 1,\\
   \frac{p}{2\Gamma(1/p)} & if \,\, \lambda =1,\\
     \frac{p(\lambda-1)^{1/p}}{2B\left(\frac{1}{p},\frac{\lambda}{\lambda-1}\right)} & if \,\, \lambda >1.
   	\end{array}\right.
 \end{equation}

 Moreover, the symbol $N_{\lambda}[G]$ denotes the Rényi entropic power (\ref{eq:entrop_gauss}) of the generalized Gaussian distribution $G(x)$ given by
    \begin{equation}
    \label{eq:entrop_gauss}
       N_{\lambda}[G] =\left[a_{p,\lambda}\,e_\lambda\left(\frac{-1}{p\lambda}\right)\right]^{-1},
      \end{equation}
and $\phi_{p,\lambda}[G]$ represents the biparametric Fisher information of the generalized Gaussian distribution \cite{lutwak} which, for $1\le p \le \infty$ and $\lambda> \frac 1 {1+p}$, is given by
     \begin{equation}
       \phi_{p,\lambda}[G]=\left\{\begin{array}{cc}
        \frac{p^{\frac1{p\lambda}}}{\lambda^{\frac1{q\lambda}}}\left[a_{p,\lambda}\,e_\lambda\left(\frac{-1}{p\lambda}\right)^{\frac1q}\right]^\frac{\lambda-1}{\lambda},& p<\infty \\  2^{(1-\lambda)/\lambda}\lambda^{\frac{-1}\lambda},& p=\infty,
         \end{array}\right.
      \end{equation}

  Note that, from (\ref{eq:def1}), (\ref{eq:renyi_entrop}) and (\ref{eq:genfish}) we can state that the biparametric complexity measure $C^{(p,\lambda)}_{FR}[\rho]$ quantifies the combined balance of the $\lambda$-dependent spreading facet of the probability distribution $\rho(x)$ and the $(p,\lambda)$-dependent oscillatory facet of $\rho(x)$. It is then clear that this quantity is much richer than e.g. the Fisher-Shannon measure which quantifies a single spreading aspect of the distribution (namely, its total spreading given by the Shannon entropy power) together with a single oscillatory facet (which corresponds to the gradient content as given by the standard Fisher information).

Indeed, the generalized complexity measure $C^{(p,\lambda)}_{FR}[\rho]$ includes the basic Fisher-Shannon complexity measure \cite{angulo_pla08,romera_1}, $C_{FS}\left[\rho\right]=\frac{1}{2 \pi e} F\left[\rho\right] \, \exp \left(2 S\left[\rho\right]\right)$, and the various one-dimensional complexity measures of Fisher-Rényi types recently published in the literature \cite{romera08,antolin09,romera09,antolin_ijqc09, rudnicki}. Most important is to point out that the novel complexity quantifier $C^{(p,\lambda)}_{FR}[\rho]$ includes the  one-parameter Fisher-Rényi complexity measure $C_{FR}^{(\lambda)}$ \cite{rudnicki}, since 
 $$C_{FR}^{(\lambda)}=(C_{FR}^{(2,\lambda)})^{2\lambda}.$$
 These two measures of complexity present a number of similarities and differences, which are worth mentioning. First, following the lines of 
\cite{rudnicki} it is straightforward to show that the biparametric measure, like the monoparametric one, has the following important properties: a universal unity lower bound ($C_{FR}^{(p,\lambda)}\ge1$), invariance under scaling and translation transformations and monotonicity The latter property, defined by Rudnicki et al [12], can be proved by use of \textit{rearrangements} (a powerful tool of functional analysis) and operating in the same manner as done in Ref. [16] for the one-parameter Fisher-R\'enyi complexity. Moreover, the biparametric measure has the following behavior under replication transformation
$$C_{FR}^ {(p,\lambda)}[\tilde \rho] =n^{\frac1\lambda}C_{FR}^ {(p,\lambda)}[\rho],$$
where the density $\tilde{\rho}$ representing $n$ replications of $\rho$ is given by
$$
\tilde{\rho}(x) = \sum_{m=1}^n\rho_m(x);\quad \rho_m(x)= n^{-\frac12} \rho\left(n^\frac12(x-b_m)\right),
$$
where the points $b_m$ are chosen such that the supports $\Delta_m$ of each density $\rho_m$ are disjoints. This property shows that this biparametric complexity quantifier, as opposed to the monoparametric one, becomes replication invariant in the limit $\lambda\to\infty$; this limiting property is an effect of the power $\frac1\lambda$ which has the biparametric measure but not the monoparametric one. In this limit, the minimizer distribution of this complexity measure has the form of a Dirac-like delta. Moreover, this power effect makes that the biparametric measure is well defined in the limit $\lambda\to\infty$, which does not happen in the monoparametric case. 

Another important difference between the bi- and uni-parametric complexity quantifiers is that the former one has two degrees of freedom; this means that it does not only depend on $\lambda$ but also on the parameter $p$. So, in particular when $\lambda=1$, we can readily show that this quantifier is minimized for Freud-like probability distributions of the form $e^{-|x|^p}$ \cite{freud,clarkson,assche}, which has a great physical relevance in the theory of sub- and super-diffusive systems.\\

There exist other instances of the biparametric complexity measure $C^{(p,\lambda)}_{FR}[\rho]$ which are relevant for different reasons. Let us just mention three of them. First, when $\lambda=1+\frac1p>\frac1{1+p}$, $\forall p>1$, we have that the resulting measure 
$$C_{FR}^{[p]}[\rho] \equiv C_{FR}^{(p,1+\frac1p)}[\rho]$$
is composed by two particularly relevant entropic factors: the generalized Fisher information 
$$\phi_{p,1+\frac1p}[\rho]=\left(\int_\Delta |\rho'(x)|^q\, dx\right)^\frac1{2q-1},$$
which is a pure functional of the derivative of the density $\rho$, and the Rényi entropic power  
$$N_{1+\frac1p}[\rho]=\left(\int_{\Delta} \left[\rho(x)\right]^{1+\frac1p}\,dx\right)^{-\frac1p}=\left\langle[\rho(x)]^{\frac1p}\right\rangle^{-p}$$
Moreover, this complexity measure is minimized by the distribution 
$$e_{p,1+\frac1p}(x)=a_{p,1+\frac1p}	\left(1-\frac{|x|^p}{p}\right)_+^p$$
Note furthermore that the support of this distribution is $[-p^\frac1p,p^\frac1p]$, which boils down to $[-1,1]$ for both values $p=1$ and $p=\infty$, and it becomes longest for $p=e$.\\

Second, when $\lambda = 2$ the corresponding complexity measure $C_{FR}^{(p,2)}[\rho]$ is proportional to the ratio $\frac{\left\langle|\rho'(x)|^q\right\rangle^\frac1{2q}}{D[\rho]}$, since the Fisher-information-factor is 
$$\phi_{p,2}[\rho]=\left(\int_\Delta \rho(x)\,|\rho'(x)|^q\, dx\right)^\frac1{2q}=\left\langle|\rho'(x)|^q\right\rangle^\frac1{2q}$$
and the Rényi-entropic-power-factor is the inverse of the disequilibrium $D(\rho)$ as
$$N_2[\rho]=\left(\int_{\Delta} \left[\rho(x)\right]^{2}\,dx\right)^{-1}=D[\rho]^{-1}$$
Moreover, the resulting measure $C_{FR}^{(p,2)}[\rho]$ is minimized by the distribution
$$e_{p,2}(x)=a_{p,2}	\left(1-|x|^p\right)_+,$$
whose support $[-1,1]$ remains invariant.
Third, most remarkable, is that the measure in the limit $p\to\infty$ is also well defined, corresponding to a step function $\forall \lambda>0.$ In the latter case the complexity measure is given by 
$$
C_{FR}^{(\infty,\lambda)}=\left(\frac\lambda2\right)^\frac1\lambda\,\phi_{\infty,\lambda}[\rho]\,N_\lambda[\rho].
$$
where the generalized Fisher-like information $\phi_{\infty,\lambda}[\rho]$ is given by
$$
\phi_{\infty,\lambda}[\rho]=\left(\int_\Delta [\rho(x)]^{\lambda-1}\,|\rho(x)'|\,dx\right)^\frac1\lambda
$$
(so that $(\phi_{\infty,\lambda}[\rho])^\lambda$ corresponds to the total variation of $\frac{\rho^\lambda}\lambda$ \cite{lutwak}) and $N_{\lambda}[\rho]$ is the previously defined Rényi entropy power. The measure $C_{FR}^{(\infty,\lambda)}$ has all the properties previously pointed out for the general biparametric Fisher-Rényi complexity. Moreover, it is minimized by the uniform distribution (as the basic LMC complexity). As well, within the set of all possible step-permutations of a generic distribution $\rho$ composed of $N$ step functions, the measure $C_{FR}^{(\infty,\lambda)}$ gets minimized by all the monotonically increasing or decreasing distributions.  Finally, let us point out that when $\frac{1}{1+p} < \lambda < 1$ the resulting complexity measures  have heavy-tailed distributions as minimizers.

\section{Application to the $d$-dimensional blackbody}

In this section, the biparametric Fisher-Rényi complexity measure $C_{FR}^{(p,\lambda)}$ is investigated for the $d$-dimensional blackbody frequency distribution at temperature $T$, $\rho(\nu) \equiv \rho_{T}^{(d)}(\nu)$, given by Eq. (\ref{eq:GPRL}). 

That is, 
\begin{equation}
\label{eq:comp_FR1}
   C_{FR}^{(p,\lambda)}[\rho_T^{(d)}] =\mathcal  K_{FR}(p,\lambda)\,\,\phi_{p,\lambda}[\rho_T^{(d)}] \times N_{\lambda}[\rho_T^{(d)}],
\end{equation}
with $ \lambda p >\frac d {d-1} $, where $\mathcal  K_{FR}(p,\lambda)$ is the normalization constant given by Eq. (\ref{eq:KFR}), and  $\phi_{p,\lambda}[\rho_T^{(d)}]$ and $N_{\lambda}[\rho_T^{(d)}]$ are the generalized Fisher information and power Rényi entropy of the $d$-dimensional blackbody density, respectively, defined in the previous section whose values will be first expressed in the following.

The Rényi entropy power is given by $N_{\lambda}[\rho_T^{(d)}] = \exp\left(R_{\lambda}[\rho_T^{(d)}] \right)$, where the Rényi entropy for the $d$-dimensional blackbody density, defined in (\ref{eq:renyi_entrop}), has been shown \cite{david_bbd} to be given by 
\begin{equation}
\label{eq:renyi_bbd} 
%\boxed{
R_{\lambda}[\rho_T^{(d)}]=\frac{1}{1-\lambda}\ln A_R(\lambda,d) + \ln \left(\frac{k _BT}{h}\right), 
\end{equation}
with $\lambda >0$ and $\lambda \neq 1$, where the constant $A_R(\lambda,d)$ has the value
\begin{equation}
\label{eq:AR}
%\boxed{
A_R(\lambda,d)=\frac{\Gamma (\lambda d+1)\zeta_\lambda(\lambda d+1,\lambda)}{\Gamma^\lambda (d+1) \zeta^\lambda(d+1)},
\end{equation}
with $\lambda\in\mathbb{N}\setminus \{1\}$, and the symbol $\zeta_{\lambda}(s,a)$ denotes the modified Riemann zeta function or Barnes zeta function \cite{barnes}.

The biparametric Fisher information $\phi_{p,\lambda}[\rho]$, defined in (\ref{eq:genfish}), for the $d$-dimensional blackbody density at temperature $T$ has been recently obtained \cite{david_bbd} as
\begin{eqnarray}
\label{eq:genfish0}
\phi_{p,\lambda}[\rho_T^{(d)}] = \left[A_F(p,\lambda,d)\right]^{\frac{1}{q\lambda}}\frac{h}{k_BT}, 
\end{eqnarray}
with $q \in (1,\infty)$,  $\lambda >0$, $\frac{1}{q} + \frac{1}{p} =1$ and where $A_F(p,\lambda,d)$ denotes the proportionality constant,
\begin{equation}
\label{eq:cteAF1}
A_F(p,\lambda,d)=\frac{I(q,\lambda,d)}{(\Gamma(d+1)\zeta(d+1))^{q\lambda -q+1}} ,
\end{equation}
 with
 \begin{equation}
 \label{eq:cteAF2}
  I(q,\lambda,d)=\int_{\mathbb{R^+}}\frac{x^{q(d\lambda -d-1)+d}}{(e^x-1)^{q\lambda +1}}\left|d(e^x-1)-xe^x\right|^q\, dx.
\end{equation}
For $d>\frac{\lambda p}{\lambda p-1}$ (so that $\phi_{p,\lambda}[\rho_T^{(d)}]$ is well-defined), $q$ even and $q\lambda \in \mathbb{N}$, the integral $I(p,\lambda,d)$ in (\ref{eq:cteAF2}) is analytically solvable giving rise to the following value for the proportionality constant

\begin{eqnarray}
\label{eq:ctefish}
A_F(p,\lambda,d) &=& (\Gamma(d+1)\,\zeta(d+1))^{-\alpha}\nonumber \\
&  &\hspace{-3cm} \times \sum_{i=0}^q(-1)^{q-i}{q\choose i}d ^i(\alpha d -i)!\,\zeta_{\alpha+q-i}(1+\alpha d -i ,\alpha),
\end{eqnarray}
%for even values of $q$ and $q\lambda \in \mathbb{N}$. 
with $\alpha \equiv q\lambda -q+1$. For the standard case $(p=2,\lambda=1)$ one obtains that
\begin{equation}
\label{eq:cteAF}
A_F(2,1,d)=\frac{1}{2\zeta(d+1)}\left(\zeta(d)-\frac{d-3}{d-1}\zeta(d-1) \right),
\end{equation}
with $d>2$.

\noindent

The insertion of (\ref{eq:renyi_bbd}) and (\ref{eq:genfish0}) into (\ref{eq:comp_FR1}) allows us to obtain finally the  expression
\begin{equation}
    C_{FR}^{(p,\lambda)}[\rho_T^{(d)}] =\mathcal K_{FR}(p,\lambda) A_F(p,\lambda,d)^{\frac{1}{q\lambda}}\,\, A_R(\lambda,d)^{\frac{1}{1-\lambda}},
\end{equation}
for the biparametric Fisher-Rényi complexity measure of the $d$-dimensional blackbody at temperature $T$, where the constants $A_R(p,\lambda,d)$ and $A_F(\lambda,d)$ are given by Eqs. (\ref{eq:AR}) and (\ref{eq:cteAF}), respectively. Note that this two-parameter complexity quantifier does not depend on $T$ nor on any physical constants (e.g., Boltzmann and Planck constants), but it does depend on the universe dimensionality only; thus, having a universal character.\\
\noindent

\begin{figure}[H]
\centering 
\subfloat[]{
\includegraphics[width=.5\linewidth]{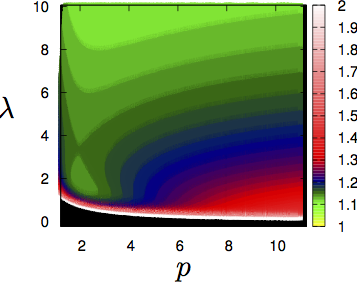}\hfill}
\caption{Representation of the Fisher-Rényi complexity for the three-dimensional blackbody distribution, $C_{FR}^{(p,\lambda)}[\rho_T^{(3)}]$, in terms of $p$ and $\lambda$.}
\label{fig:fig1}
\end{figure}

\begin{figure}[H]
\centering
  \subfloat[]{%
  \begin{minipage}{\linewidth}
  \includegraphics[width=.5\linewidth]{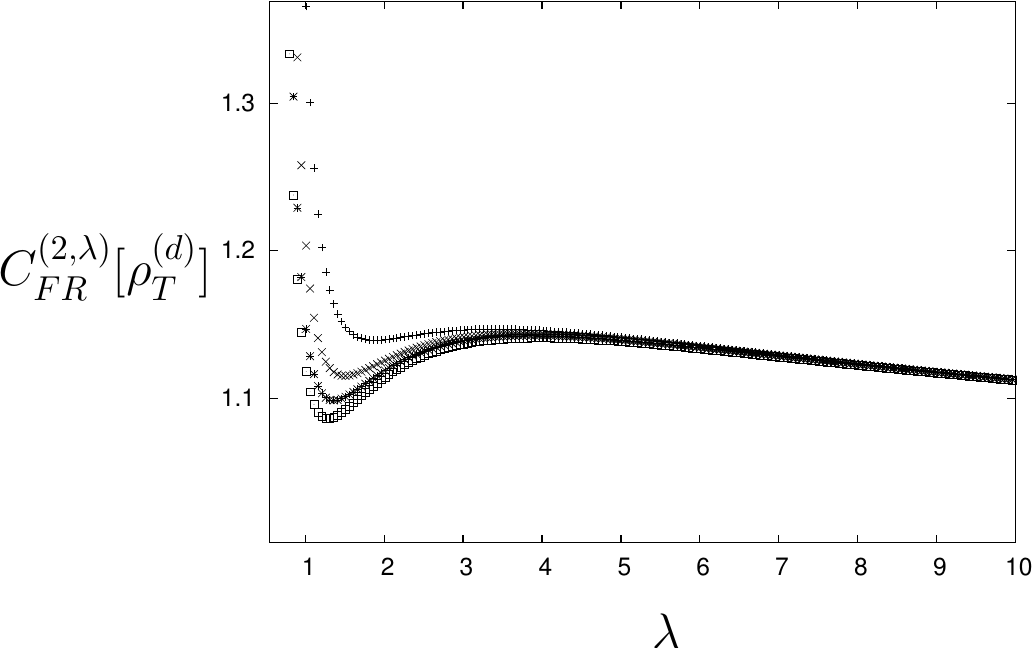}\hfill
  \includegraphics[width=.5\linewidth]{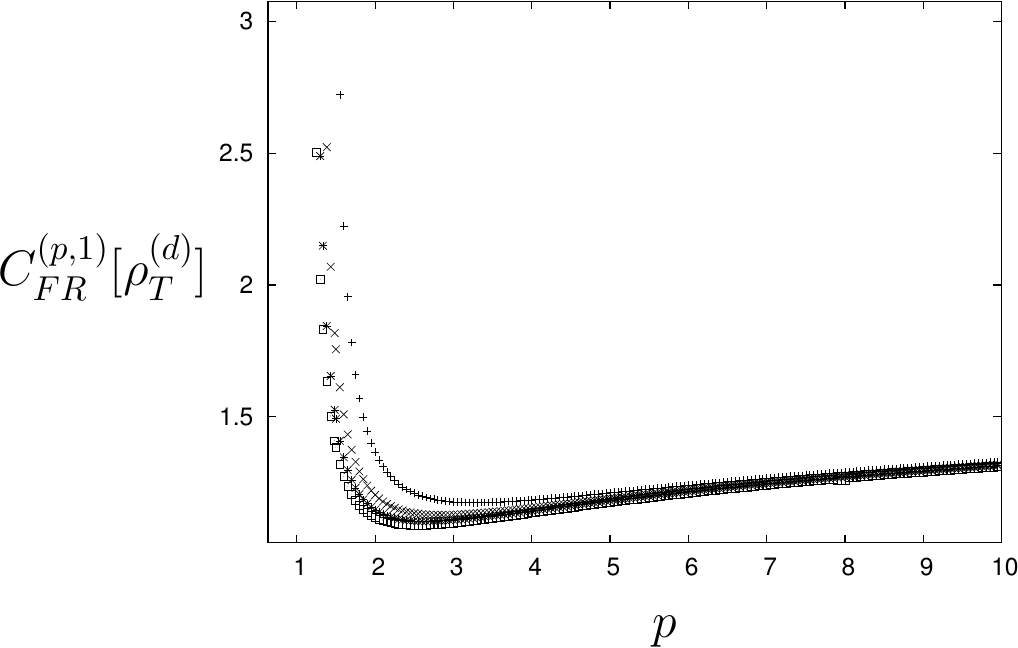}\hfill
  \end{minipage}%
  }\par
 \caption{Left: Dependence of the $d$-dimensional Fisher-Rényi complexity $C_{FR}^{(2,\lambda)}[\rho_T^{(d)}]$ on $\lambda$ when $d= 3(+), 4(\times),5 (*), 6(\square)$. Right: Dependence of the $d$-dimensional Fisher-Rényi complexity $C_{FR}^{(p,1)}[\rho_T^{(d)}]$ on $p$ when $d= 3(+), 4(\times),5 (*), 6(\square)$.}
 \label{fig:fig2}
\end{figure}

For a better understanding of how the biparametric Fisher-Rényi complexity measure $C_{FR}^{(p,\lambda)}[\rho_T^{(d)}]$ is able to characterize the multidimensional blackbody distribution, we study its dependence on the spatial dimensionality $d$ and the parameters $p$ and $\lambda$ in Figures 1 and 2. In Fig. \ref{fig:fig1} we represent the $(p,\lambda)$-chart of the Fisher-Rényi complexity for the three-dimensional blackbody distribution, $C_{FR}^{(p,\lambda)}[\rho_T^{(3)}]$, in terms of $p$ and $\lambda$. This quantity has no physical units and it does not depend on the blackbody temperature, what highlights the universal character of the chart. We observe that the $(p,\lambda)$-Fisher-Rényi complexity (i) presents a relative minimum at around $(p=2.20,\lambda=1.74)$, (ii) for all fixed $\lambda$ it has  an absolute minimum whose location depends on $\lambda$, appearing a complexity horizontal asymptote at values bigger than unity, and opposite (iii) for fixed $p$ it has the three following regimes. For $1\le p \lesssim 1.55$ the $\lambda$-behavior of the complexity is monotonically decreasing and strictly concave; for $ 1.55\lesssim p \lesssim 4.3$ this behavior breaks down, appearing two critical points (see also Figure \ref{fig:fig2} left), and finally for $ 4.3 \lesssim p$ the critical points dissapear and the concave behaviour go back slowly when $p$ is increasing. Nevertheless when $\lambda$ is large enough the complexity goes to unity independently of the $p$-value.  These phenomena can be also observed in the functions $C_{FR}^{(2,\lambda)}[\rho_T^{(3)}]$ and $C_{FR}^{(p,1)}[\rho_T^{(3)}]$ of Fig. \ref{fig:fig2}, which corresponds to the cuts at $p=2$ y $\lambda=1$ of the $(p,\lambda)$-chart, respectively.  

Besides we study in Fig. 2 the dimensionality dependence of the Fisher-Rényi complexity measures $C_{FR}^{(2,\lambda)}[\rho_T^{(d)}]$ and $C_{FR}^{(p,1)}[\rho_T^{(d)}]$ when $d= 3(+), 4(\times),5 (*), 6(\square)$ in terms of the corresponding parameters $\lambda$ and $p$, respectively. Note that the dimensionality behavior is qualitatively similar in each case. The complexity $C_{FR}^{(2,\lambda)}[\rho_T^{(d)}]$ as a function of $\lambda$ has, for all dimensionalities, a minimum and a maximum within the interval $(0,4)$ and then it monotonically decreases to unity. On the opposite, the complexity $C_{FR}^{(p,1)}[\rho_T^{(d)}]$ as a function of $p$ has only a minimum at $p<4$ for all dimensionalities and then it monotonically grows when $p$ is increasing. In both cases the minimum location decreases when the dimensionality is augmenting. Quantitatively, we observe that for $\lambda \geq 5$ in the left graph and for $p\geq 4$ in the right graph the corresponding complexities do not practically depend on the dimensionality.

\section{Conclusions and open problems}

First we have shown in this paper that the Rényi entropy power, $N_{\lambda}[\rho]$, (that generalizes the Shannon entropy power) and the biparametric Fisher information, $\phi_{p,\lambda}[\rho]$, (which generalizes the standard Fisher information) allow us to construct a novel class of generalized complexity measures for a general probability density $\rho(x)$, the biparametric Fisher-Rényi complexities $C^{(p,\lambda)}_{FR}[\rho]$. They quantify jointly the $\lambda$-dependent spreading aspects and the $(p,\lambda)$-dependent oscillatory facets of $\rho(x)$, so being much richer than the basic Fisher-Shannon measure and all its extensions of Fisher-Rényi type. Second, we have pointed out a number of properties of these quantifiers, such as universal lower bound, scaling and translation invariance and monotonicity, among others.\\

Third, we have applied these complexity quantifiers to the $d$-dimensional blackbody radiation distribution at temperature $T$. We have found that they do not depend on the temperature nor on any physical constant (Planck constant, speed of light and Boltzmann constant) but only on the spatial dimensionality, which gives them a universal character. We are aware that the full power of the novel complexity quantifiers here proposed will be shown in multimodal/anisotropic probability distributions so abundant in natural phenomena, such as e.g. the ones which are being observed in the present observational missions of the various cosmic frequency backgrounds of the universe from the known microwave and infrared ones to the emergent cosmic neutrino and gravitational backgrounds; this is because in such cases this novel class of quantifiers may be used to detect and quantify the inherent anisotropies because of the sensitivity of its generalized-Fisher-information factor which captures the density fluctuations in a multi-faceted manner. Moreover, the application of this class of complexity quantifiers to fractal phenomena as well as to the non-linear blackbody radiation laws \cite{zeng,plastino,tsallis,valluri}, which presumably takes into account the small deviations from the Planck radiation formula that have been recently detected in the cosmic microwave radiation, are two relevant problems which deserve to be explored.\\

\section*{Acknowledgments}
This work was partially supported by the grants FQM-7276 and FQM-207 of the Junta de Andaluc\'ia and the MINECO-FEDER (Ministerio de Economía y Competitividad, and the European Regional Development Fund) grants FIS2014-54497P and FIS2014-59311P. The work of I. V. Toranzo was financed by the program FPU of the Spanish Ministerio de Educación.

\end{document}